\documentstyle[12pt,aps,preprint]{revtex}

\def\W#1{\ ^{W}\!\!#1}
\def\cris#1#2{\left\{{#1\atop #2} \right \}}
\draft
\title{\flushright{\small \sc AEI--2002--060}\\
\centerline{Weyl Geometry and Quantum Gravity}}
\author{FATIMAH SHOJAI\thanks{Email: FATIMAH@IPM.IR}}
\address{Institute for Studies in Theoretical Physics and Mathematics, P.O.Box 19395-5531, Tehran,
IRAN}
\address{and}
\address{MPI f.  Gravitationsphysik, Albert--Einstein--Institut, Am M\"uhlenberg
1, 14476  Golm  near  Potsdam,  Germany}
\author{ALI SHOJAI\thanks{Email: SHOJAI@IPM.IR}}
\address{Physics Department, Tehran University, End of North karegar St., Tehran 14352, IRAN}
\address{and}
\address{Institute for Studies in Theoretical Physics and Mathematics, P.O.Box 19395-5531, Tehran,
IRAN}
\address{and}
\address{MPI f.  Gravitationsphysik, Albert--Einstein--Institut, Am M\"uhlenberg
1, 14476  Golm  near  Potsdam,  Germany}
\begin{document}
\maketitle
\begin{abstract}
It is shown
that the recently geometric formulation of quantum
mechanics\cite{GEO,ST,CT,NL,QGG,SST} implies the use of Weyl
geometry. It is discussed that the natural framework for both
gravity and quantum is Weyl geometry. At the end a Weyl invariant
theory is built, and it is shown that both gravity and quantum are
present at the level of equations of motion. The theory is applied 
to cosmology leading to the desired time dependencies of the cosmological
and gravitational constants.
\end{abstract}
\section{INTRODUCTION AND SURVEY}
Why Weyl geometry? From a long time ago it is believed that the
long range forces (i.e. electromagnetism and gravity) are
different aspects of a unique phenomena. So they must be unified.
Usually it is proposed that one must generalize Einstein's
general relativity theory to have a geometrical description of
electromagnetic fields. This means to change the properties of the
manifold of general relativity. Using higher dimensional
manifolds\cite{Kaluza,Over}, changing the compatibility relation
between the metric and the affine connection\cite{Weyl,Adler} and using
a non-symmetric metric\cite{Ein,Sch} are some examples of the
attempts towards this idea. In all the above approaches, the
additional degrees of freedom correspond  to the components of the
electromagnetic potential.

The second idea leads to the Weyl's gauge invariant
geometry. In Weyl geometry, both the components and the length of
a vector change linearly proportional to the infinitesimal
translation during any parallel transportation. The former
produces some rotation of the vector, as in Riemanian geometry,
while the latter  is a special aspect of Weyl geometry expressed
as:
\begin{equation}
\delta\ell=\phi_\mu\delta x^\mu\ell ; \ \ \ \ or\ \ \ell=\ell_0 \exp \left(\int 
\phi_\mu dx^\mu\right)
\label{x1}
\end{equation}
So the length unit is different at different points. Relation
(\ref{x1}) presents a new vector field ($\phi_\mu$) in the
theory, identified with the electromagnetic potential by
Weyl\cite{Weyl,Adler}. This equation shows that the change of length
between two arbitrary points is dependent on the chosen path
unless the curl of the vector field is zero (non--integrability of lenght). 

Equivalently, the length change can be replaced by a change in the
metric as:
\begin{equation}
g_{\mu\nu}\rightarrow \exp\left(2\int 
\phi_\mu dx^\mu\right)g_{\mu\nu} 
\end{equation}
which is called Weyl transformation.
We see that the metric is a Weyl covariant (or co-covariant) object of the weight 2. 
Now  assumming that contravariant vectors change during any parrallel transportation 
as in Riemanian geometry, relation (\ref{x1}) shows that    
the Weyl affine connection is given
by:
\begin{equation}
\W{\Gamma}^\mu_{\nu\lambda}=\cris{\mu}{\nu\lambda}+g_{\nu\lambda}\phi^\mu
-\delta^\mu_\nu\phi_\lambda-\delta^\mu_\lambda\phi_\nu
\end{equation}
differring from Christoffel symbols by the last
three terms.

Now suppose that we make the following transformation:
\begin{equation}
\phi_\mu\longrightarrow\phi'_\mu=\phi_\mu+\partial_\mu\Lambda
\label{x2}
\end{equation}
which is called a {\it gauge transformation\/}. The effect of this is to transform 
$g_{\mu\nu}\rightarrow g'_{\mu\nu}=\exp(2\Lambda)g_{\mu\nu}$ and 
$\delta\ell\rightarrow \delta\ell ' =\delta\ell +(\partial_\mu\Lambda)\delta x^\mu \ell$. 
One can see the change in lenght is the same in the two gauges when one turns around 
a closed path. So a gauge transformation does not change the geometry.

The quantity defined as the curl of the Weyl vector:
\begin{equation}
F_{\mu\nu}=\partial_\mu\phi_\nu-\partial_\nu\phi_\mu
\end{equation}
is gauge invariant, and corresponds to the electromagnetic field. If the Weyl 
vector be the gradient of a
scalar, there exists a gauge transformation leading to a zero Weyl
vector field or equivalently Riemannian geometry class. In this case the lenght 
is integrable. So the properties of the elementary particles are independent of 
their path history. 

Appart from the electromagnetic aspects of Weyl geometry, it has some other applications.
Some authors believe that Weyl geometry is a suitable framework for quantum gravity.
In ref\cite{Wheeler} a new quantum theory is
proposed on the basis of Weyl picture which is purely geometric.
The observables are introduced as zero Weyl weight quantities.
Moreover any weightful field has a Weyl conjugate such as complex
conjugate of the state vector in quantum mechanics. By these dual
fields, the probability can be defined. These are the elements of
a consistent quantum theory which is equivalent to the standard
quantum mechanics. Moreover it is shown that the quantum
measurement and the related uncertainty  would emerged from Weyl
geometry naturally. In this theory when the curl of Weyl vector is
zero, we arrive at the classical limit. By noting the relation(\ref{x2}),
it is concluded that the change of length scale is only a quantum
effect.

Another different approach to geometrize quantum mechanics can be found in 
\cite{Wood}. Here a modified Weyl--Dirac theory is used to join the particle 
aspects of matter and Weyl symmetry breaking.

In the present
work we shall look at the conformal invariance at the quantum
level. Does the quantum theory lead us to any characteristic
length scale and thus break the conformal symmetry? Or conversely
the quantum effects lead us to a conformal invariant geometry? We
shall discuss these questions in the context of the causal quantum
theory proposed by Bohm\cite{Bohm1,Bohm2,Hi}.

This paper is organized in the following manner. 
In section II we shall discuss the Weyl--Dirac theory in
details. One of the main points of this paper comes in section III
where we shall show that the Weyl vector and the quantum
effects of matter are connected, so answering the question why Weyl geometry. In
this sense it may be suitable to name Weyl geometry as quantum
geometry. We shall preciesly show in this section how the conformal symmetry 
emerges naturally by considering quantum effects of matter. Finally in section 
IV we show that the Weyl--Dirac theory is a suitable framwork for 
identification of  the conformal degree of
freedom of the space--time with the Bohm's quantum mass.

\section{Weyl--Dirac theory}
Straightforward generalization of Einstein--Hilbert action to Weyl geometry leads 
to a higher order theory\cite{Weyl,Adler}. Dirac\cite{Dirac,Ros} introduced 
a new action called Weyl--Dirac action, by including a new field which is in 
fact gauge function. It helps him
to avoid higher order actions as while as fixing the gauge function leads to 
Einstein--Maxwell equations.

The Weyl--Dirac action is given by\cite{Dirac,Ros}:
\begin{equation}
{\cal A}=\int d^4x \sqrt{-g} \left ( F_{\mu\nu}F^{\mu\nu}-\beta^2\W{{\cal R}}+(\sigma +6)
\beta_{;\mu}\beta^{;\mu}+{\cal L}_{matter}\right )
\end{equation}
where $\beta$ is a scalar field of weight $-1$. The ``$;$''
represents covariant derivation under general coordinate and
conformal transformations (Weyl covariant derivative) defined as:
\begin{equation}
X_{;\mu}=\W{\nabla}_\mu X-{\cal N}\phi_\mu X
\end{equation}
where ${\cal N}$ is the Weyl weight of $X$.
The equations of motion then would be:
\[
{\cal G}^{\mu\nu}=-\frac{8\pi}{\beta^2}({\cal T}^{\mu\nu}+M^{\mu\nu})+\frac{2}{\beta}(
g^{\mu\nu}\W{\nabla}^\alpha\W{\nabla}_\alpha\beta-\W{\nabla}^\mu\W{\nabla}^\nu\beta)
\]
\begin{equation}
 +\frac{1}{\beta^2}(4\nabla^\mu\beta\nabla^\nu\beta-g^{\mu\nu}\nabla^\alpha\beta \nabla_\alpha\beta) 
+\frac{\sigma}{\beta^2}(\beta^{;\mu}\beta^{;\nu}-\frac{1}{2}g^{\mu\nu}\beta^{;\alpha} \beta_{;\alpha})
\end{equation}
\begin{equation}
\W{\nabla}_{\nu}F^{\mu\nu}=\frac{1}{2}\sigma(\beta^2\phi^\mu+\beta\nabla^\mu\beta) +4\pi J^\mu
\end{equation}
\begin{equation}
{\cal R}=-(\sigma+6)\frac{\W{\Box}\beta}{\beta}+\sigma\phi_\alpha\phi^\alpha 
-\sigma \W{\nabla}^\alpha\phi_\alpha+\frac{\psi}{2\beta}
\end{equation}
where:
\begin{equation}
M^{\mu\nu}=\frac{1}{4\pi}\left ( \frac{1}{4}g^{\mu\nu}F^{\alpha\beta}F_{\alpha \beta} -F^\mu_\alpha F^{\nu\alpha}\right )
\end{equation}
and the energy--momentum tensor ${\cal T}^{\mu\nu}$, current density vector $J^\mu$ and the scalar $\psi$ are defined as:
\begin{equation}
8\pi{\cal T}^{\mu\nu}=\frac{1}{\sqrt{-g}}\frac{\delta \sqrt{-g}{\cal L}_{matter}}{\delta g_{\mu\nu}}
\end{equation}
\begin{equation}
16\pi J^\mu=\frac{\delta {\cal L}_{matter}}{\delta \phi_\mu}
\end{equation}
\begin{equation}
\psi=\frac{\delta {\cal L}_{matter}}{\delta \beta}
\end{equation}
On the other hand the equation of motion of matter and the trace 
of energy-momentum tensor can be resulted from the invariance of 
action under the coordinate and gauge transformations. One can 
write them as respectively:
\begin{equation}
\W{\nabla}_\nu{\cal T}^{\mu\nu}-{\cal T}\frac{\nabla^\mu\beta}{\beta}=J_\alpha\phi^{\alpha\mu} 
- (\phi^\mu+\frac{\nabla^\mu\beta}{\beta})\W{\nabla}_\alpha J^\alpha
\end{equation}
\begin{equation}
16\pi{\cal T}-16\pi\W{\nabla}_\mu J^\mu-\beta\psi=0
\end{equation}
The first of them is only a geometrical identity (Bianchi identity) and the 
second results from the non independence of field equations.

It must be noted that in the Weyl--Dirac theory, the Weyl vector does not couple 
to spinors, so $\phi_\mu$ cannot be interpretad as the electromagnetic potential\cite{Lord}.
Here we use the Weyl vector not as the electromagnetic field but only as a part 
of the geometry of the space--time. The Weyl--Dirac formalism is adopted and we shall 
see that  the auxiliary field (gauge function) in Dirac's action represents the 
quantum mass field. In addition both gravitation fields ($g_{\mu\nu}$ and $\phi_\mu$) 
and quantum mass field determine the geometry of the space--time. 

\section{Bohmian quantum gravity and Weyl symmetry}
In a series of papers\cite{GEO,ST,CT,NL,QGG,SST} we have proposed
a new approach to quantum gravity based on the de-Broglie--Bohm
quantum theory. The approach in the above
references is different from the standard Bohmian quantum gravity
presented e.g. in \cite{Holland}. These works are attempts to
geometrize the quantum behaviour of matter.
This point, as a conjecture, firstly proposed by
de-Broglie\cite{de} which states that the quantum effects
can be removed via a conformal transformation. We have shown in
\cite{GEO} that the quantum effects can be included in the
conformal degree of freedom of the  space--time metric. Adding the
gravitational effects, it can be seen that quantum and gravity are
highly coupled. This produces remarkable changes in the classical
predictions such as the physics of the birth of the universe
as long as the classical limit is obtained correctly.

One of the new points of the above approach is the dual role of
geometry in physics. The gravitational effects determine the
causal structure of the space--time as long as the quantum effects
give its conformal structure. This does not mean that the quantum
effects has nothing to do with the causal structure, it can act on
the causal structure through back--reaction terms appeared in the
metric field equations\cite{ST,CT,SST}. We only mean that the
dominant term in the causal structure is the gravitational
effects. The same is true for the conformal factor. The conformal
factor of metric is a function of quantum potential which is the
principal character in Bohm's theory and given by:
\begin{equation}
{\cal Q}=\alpha\frac{\Box\sqrt{\rho}}{\sqrt{\rho}}; \ \ \ \ \ \ \ \alpha=\frac{\hbar^2}{m^2c^2}
\end{equation}
where $\rho$ is the ensemble density of the system.
According to Bohm the mass of a relativistic particle is a field
produced by quantum corrections to classical mass. This can be
easily seen from the quantum-Hamilton-Jacobi equation for a
spin-less particle:
\begin{equation}
\nabla_\mu S\nabla^\mu S={\cal M}^2c^2=m^2c^2(1+{\cal Q})
\label{x11}
\end{equation}
where $S$ is the Hamilton--Jacobi function and ${\cal M}$ is the quantum mass.
We have shown in ref\cite{ST,CT,SST} that the presence of quantum
potential is equivalent to a conformal mapping of the metric. Thus
in the conformally related frames we feel different quantum masses
and different curvatures correspondingly. It is possible to
consider two specific frames. One of them contains the quantum
mass field (appeared in quantum Hamilton-Jacobi equation) and the
classical metric as long as in the other the classical mass
(appeared in classical Hamilton-Jacobi equation) and quantum
metric are appeared. In other frames both the space--time
metric and mass field contain the quantum properties. This
argument motivates us to state that {\it different conformal
frames are identical pictures  of the gravitational and quantum
phenomena\/}. Considering the quantum force, the conformally
related frames aren't distinguishable. This is just what happens
when we consider gravity, different coordinate systems are
equivalent. Since the conformal transformation change the length
scale locally, therefore we feel different quantum forces in different 
conformal frames. This is similar to general relativity in which general
coordinate transformation changes the gravitational force at any
arbitrary point. Here it may be appropriate to state a basic
question. Does applying the above correspondence, between quantum and
gravitational forces, and between the conformal and general
coordinate transformations, means that the geometrization  of
quantum effects implies the conformal invariance  as
gravitational effects imply the general coordinate invariance?

In order to discuss this question, we  recall  what were
considered early in the development of the theory of general
relativity. General covariance principle leads to the
identification of gravitational effects of matter with the
geometry of the space--time. In general relativity the important
fact which supports this identification is the equivalence
principle. According to it, one can always remove the
gravitational field at some point by a suitable coordinate
transformation. Similarly, as we pointed out previously, according to our new
approach to Bohmian quantum gravity, at any point (or even
globally) the quantum effects of matter can be removed by a
suitable conformal transformation. Thus in that point(s) matter
behaves classically. In this way we can introduce a new
equivalence principle calling it as {\it conformal equivalence
principle\/} similar to the standard equivalence
principle\cite{QGG}. The latter interconnects gravity and general
covariance while the former has the same role about quantum
and conformal covariance. Both of these principles state that
there isn't any preferred frame, either coordinate or conformal
frame. Since Weyl geometry welcomes conformal invariance and since it 
has additional degrees of freedom which can be identified with quantum effects, it
provides a unified geometrical framework for
understanding the gravitational and quantum forces. In this way a pure geometric
interpretation of quantum behavior can be built.

Because of these results, we believe  that the de-Broglie--Bohm
theory must receive increasing attention in quantum gravity. This
theory has  some important features. One of them is that the
quantum effects appear independent of any preferred scale length (this
is  opposite to the standard quantum mechanics in which Plank
length is the characteristic  length). This is one of the
intrinsic properties of this theory which resulted from the
special definition of the classical limit\cite{Holland}. Another
important aspect is that the quantum mass of the particle is a
field. This is needed for having conformal invariance, since mass
has a non--zero Weyl weight. Also as we have shown
previously\cite{ST,CT,SST} the guiding equation lead us to the
following geodesic equation:
\begin{equation}
\frac{d^2x^\mu}{d\tau^2}+\cris{\mu}{\nu\lambda}
\frac{dx^\nu}{d\tau}\frac{dx^\lambda}{d\tau}=\frac{1}{{\cal M}}
\left ( g^{\mu\nu}-\frac{dx^\mu}{d\tau}\frac{dx^\nu}{d\tau}\right
)\nabla_\nu{\cal M}
\label{x12}
\end{equation}
The appearance of quantum mass justifies the Mach's principle\cite{Demaret}
which leads to the existence of interrelation between global
properties of the universe (space--time structure, the large scale
structure of the universe,$\cdots$) and its local properties
(local curvature, motion in a local frame,$\cdots$). In the
present theory, it can be easily seen that the geometry of the
space--time is determined by the distribution of matter. A local
variation of matter field distribution changes the quantum
potential acting on geometry. Thus the geometry would be altered
globally (in conformation with Mach's principle). In this sense
our approach to the quantum gravity is highly non--local as it is
forced by the nature of the quantum potential\cite{NL}. What we
call geometry is only the gravitational and quantum effects of
matter. Without matter the geometry would be meaningless. Moreover
in \cite{ST,SST} we have shown that it is necessary to assume an
interaction term between the cosmological constant (large scale
structure) and the quantum potential (local phenomena). These
properties all justify Mach's principle. It is shown in
\cite{ST,SST} that the gravitational constant is in fact a field
depending on the matter distribution through quantum potential.

All these arguments based on Bohmian quantum mechanics motivates
us that the Weyl geometry is a suitable framework for formulating
quantum gravity.

\section{WEYL INVARIANT QUANTUM GRAVITY}
In this section we shall construct a theory for Bohmian quantum
gravity which is conformal invariant in the framework of Weyl
geometry. To begin, note that if our model should consider massive
particles, the mass must be a field. This is because mass has
non--zero Weyl weight. This is in agreement with Bohm's theory. 
As we argued previousely a general Weyl invariant action is the 
Weyl--Dirac action, whose equations of motion are derived in 
section II. To simplify our model, we assume that the matter lagrangian 
does not depends on the Weyl vector, so that $J_\mu=0$. The equations of motion are now:
\[
{\cal G}^{\mu\nu}=-\frac{8\pi}{\beta^2}({\cal T}^{\mu\nu}+M^{\mu\nu})+\frac{2}{\beta}(
g^{\mu\nu}\W{\nabla}^\alpha\W{\nabla}_\alpha\beta-\W{\nabla}^\mu\W{\nabla}^\nu\beta)
\]
\begin{equation}
 +\frac{1}{\beta^2}(4\nabla^\mu\beta\nabla^\nu\beta-g^{\mu\nu}\nabla^\alpha\beta 
\nabla_\alpha\beta) +\frac{\sigma}{\beta^2}(\beta^{;\mu}\beta^{;\nu}-\frac{1}{2}g^{\mu\nu}\beta^{;\alpha} 
\beta_{;\alpha})
\end{equation}
\begin{equation}
\W{\nabla}_{\nu}F^{\mu\nu}=\frac{1}{2}\sigma(\beta^2\phi^\mu+\beta\nabla^\mu\beta) 
\label{x3}
\end{equation}
\begin{equation}
{\cal R}=-(\sigma+6)\frac{\W{\Box}\beta}{\beta}+\sigma\phi_\alpha\phi^\alpha 
-\sigma \W{\nabla}^\alpha\phi_\alpha+\frac{\psi}{2\beta}
\label{x4}
\end{equation}
and symmetry conditions are:
\begin{equation}
\W{\nabla}_\nu{\cal T}^{\mu\nu}-{\cal T}\frac{\nabla^\mu\beta}{\beta}=0
\label{x6}
\end{equation}
\begin{equation}
16\pi{\cal T}-\beta\psi=0
\label{x5}
\end{equation}
It must be noted that from equation (\ref{x3}) we have:
\begin{equation}
\W{\nabla}_\mu \left ( \beta^2\phi^\mu+\beta\nabla^\mu\beta\right )=0
\label{x13}
\end{equation}
so $\phi_\mu$ is not independent of $\beta$.

It is worthwhile to see whether this model has anything to do with
the Bohmian quantum gravity or not. We want to introduce the quantum mass field. 
Now we    shall show that this field is proportional to the Dirac field. In 
order to see this two conditions are neccessary to meet. Firstly the correct 
dependence of Dirac field on the trace of energy--momentum tensor and secondly 
the correct appearance of quantum force in the geodesic equation. Now note that using
equations (\ref{x3}),(\ref{x4}), and (\ref{x5}) we have:
\begin{equation}
\Box\beta+\frac{1}{6}\beta{\cal R}=\frac{4\pi}{3}\frac{{\cal T}}{\beta}+\sigma\beta 
\phi_\alpha\phi^\alpha +2 ( \sigma-6)\phi^\gamma\nabla_\gamma \beta 
+\frac{\sigma}{\beta} \nabla^\mu\beta\nabla_\mu\beta
\end{equation}
This equation can be solved iteratively.
Let we rewrite it as:
\begin{equation}
\beta^2=\frac{8\pi{\cal T}}{{\cal R}}-\frac{1}{{\cal R}/6-\sigma\phi_\alpha\phi^\alpha} 
\beta\Box\beta +\cdots
\end{equation}
The first and the second order solutions of this equation is:
\begin{equation}
\beta^{2(1)}=\frac{8\pi{\cal T}}{{\cal R}}
\end{equation}
\begin{equation}
\beta^{2(2)}=\frac{8\pi{\cal T}}{{\cal R}}\left ( 1-\frac{1}{{\cal R}/6-\sigma\phi_\alpha\phi^\alpha} 
\frac{\Box\sqrt{{\cal T}}}{\sqrt{{\cal T}}}+\cdots \right )
\label{x9}
\end{equation}
In order to derive the geodesic equation we use the relation (\ref{x6}). 
Assuming that matter is consisted of dust
with the energy--momentum tensor:
\begin{equation}
{\cal T}^{\mu\nu}=\rho u^\mu u^\nu
\label{x7}
\end{equation}
where $\rho$ and $u^\mu$ are matter density and velocity
respectively, substituting (\ref{x7}) into (\ref{x6}) and multiplying by $u_\mu$, gives us:
\begin{equation}
\W{\nabla}_\nu(\rho u^\nu)-\rho\frac{u_\mu\nabla^\mu\beta}{\beta}=0
\label{x8}
\end{equation}
If we substitute (\ref{x6}) into (\ref{x8}) again, we have:
\begin{equation}
u^\nu\W{\nabla}_\nu u^\mu=\frac{1}{\beta}(g^{\mu\nu} -u^\mu u^\nu ) \nabla_\nu\beta
\label{x10}
\end{equation}
Comparison of equations (\ref{x9}) and (\ref{x10}) with equations (\ref{x11}) and (\ref{x12}) 
shows that we have the coorect equations of motions of Bohmian quantum gravity, provided we identify:
\begin{equation}
\beta\longrightarrow {\cal M}
\end{equation}
\begin{equation}
\frac{8\pi{\cal T}}{{\cal R}}\longrightarrow m^2
\end{equation}
\begin{equation}
\frac{1}{\sigma\phi_\alpha\phi^\alpha-{\cal R}/6}\longrightarrow \alpha
\end{equation}

\section{APPLICATION TO COSMOLOGY}
Most of physicists believe in a non--zero cosmological constant because of two important reasons.
It helps us to make the theoretical results to agree with observations. Morever some topics, like 
large scale structure of the universe, dark matter, inflation, can be explored using it. On the 
other hand from astronomical observations, especially gravitational lensing, cosmological constant 
should be very small. ($|\Lambda|<10^{-54}/cm^2$) The fact that the cosmological constant is small
produces some difficulties. How explain theoretically this value of the cosmological constant? (This 
question also applies to the gravitation coupling constant.) Morever the cosmological constant is a
measure of vaccum energy density. This includes some contribution from scalar fields, bare cosmological
constant, quantum effect, and so on. But observed cosmological constant is more smaller than (120 order of
magnitude less than) each one of the above contributions. This is the so--called cosmological constant
puzzle (see \cite{CAR} and its references). Till now many mechanisms are presented to solve the problem. 

One way to solve the problem is to give dynamical characters to gravitational and cosmological constants
in such a way that they decrease as the universe expands. 
Some works are done in \cite{Hor} and \cite{Sal}. In the former, a mechanism is presented using the
WDW equation, while the latter, focuses on the breaking the conformal invariance. Two scales, cosmological
and particle physics are introduced. And a dynamical conformal factor which relates them produces an effective
time dependent cosmological constant. 

We also use the conformal invariance, but in the conformal invariant framework 
of the present paper. Let's choose a spatially flat
Robertson--Walker metric:
\begin{equation}
ds^2=a^2(\eta)\left [ d\eta^2-dr^2-r^2d\Omega^2\right ]
\end{equation}
where $a(\eta)$ is the scale factor, and assuming the universe is filled of a dust, 
the equations of motion of theory presented in the previous section now simplifies to:
\begin{equation}
3\frac{\dot{a}^2}{a^4}-\frac{8\pi\rho}{\beta^2}+\frac{6}{\beta}\left ( \frac{\dot{a}}{a}-\phi\right ) \frac{\dot{\beta}}{a^2}+
\frac{3}{\beta^2}\frac{\dot{\beta}^2}{a^2} +\frac{\sigma}{2\beta^2}\frac{(\dot{\beta}+\phi\beta)^2}{a^2}=0
\end{equation}
\begin{equation}
\dot{\beta}+\beta\phi=0
\end{equation}
\begin{equation}
-6\frac{\ddot{a}}{a^3}-(\sigma+6)\left ( \frac{1}{\beta}\frac{d}{d\eta}\left (\frac{\dot{\beta}}{a^2}\right ) +\frac{\dot{\beta}}{\beta a^2}
\left (4\frac{\dot{a}}{a}-10\phi\right ) \right ) +\sigma \frac{\phi^2}{a^2}-\sigma \frac{d}{d\eta}\left (\frac{\phi}{a^2}
\right ) -\sigma \frac{\phi}{a^2}
\left ( 4\frac{\dot{a}}{a}-10\phi\right )+\frac{\psi}{2\beta}=0
\end{equation}
where a dot over any quantity represents derivation with respect to time and we have chosen the gauge
\begin{equation}
\phi_\mu=(\phi,0,0,0)
\end{equation}
And the symmetry conditions are:
\begin{equation}
\dot{\rho}+3\rho\left ( \frac{\dot{a}}{a}-\phi\right )-\rho\frac{\dot{\beta}}{\beta}=0
\end{equation}
\begin{equation}
16\phi\rho-\beta\psi=0
\end{equation}
Introducing the cosmological time as $dt=ad\eta$ and simplifying the relations, we finally have:
\begin{equation}
\rho a^3\beta^2=constant
\end{equation}
\begin{equation}
3\frac{a'^2}{a^2}-\Lambda_{eff}-8\pi G_{eff}\rho=0
\end{equation}
\begin{equation}
3\frac{a''}{a}+3\frac{a'^2}{a^2}+30\frac{\beta'^2}{\beta^2}+9\frac{a'}{a}\frac{\beta'}{\beta}+3\frac{\beta''}{\beta}-4\pi G_{eff}\rho=0
\end{equation}
where a $'$ over any quantity represents derivation with respect to the cosmological time and we have deffined:
\begin{equation}
\Lambda_{eff}=-9\frac{\beta'^2}{\beta^2}-6\frac{a'}{a}\frac{\beta'}{\beta}
\end{equation}
\begin{equation}
G_{eff}=\frac{1}{\beta^2}
\end{equation}
The above equations can simply solved resulting in:
\begin{equation}
H\sim t^{-1}
\end{equation}
\begin{equation}
\Lambda_{eff}\sim t^{-2}
\end{equation}
\begin{equation}
G_{eff}\sim t^{-4/19}
\end{equation}
where $H$ is the Hubble constant. As the universe expands these quantities 
decrease in agreement with the above disscusion. These constants 
have a small value at the current epoch as the observation suggests.
\section{CONCLUSION}
We addressed the question ``{\it Why Weyl geometry?\/}". Among all
the arguments in the favor of it, the most important one is that
the conformal degree of freedom of the space--time metric, should
be identified with Bohm's quantum potential. We saw that one can
formulate a {\it generalized equivalence principle\/} which states
that gravitation can be removed locally via an appropriate
coordinate transformation, while quantum force can be removed
either locally or globally via an appropriate scale
transformation. So the natural framework of quantum and gravity is
Weyl geometry. The most
simplest Weyl invariant action functional is written out. It
surprisingly leads to the correct Bohm's equations of motion. When
it applied to cosmology it leads to time decreasing cosmological and gravitational constants.
A phenomena which is good for describing their small values.

Since a gauge transformation can transform a general space--time dependent Dirac field to
a constant one, and vice-versa, it can be shown that quantum effects and the lenght 
scale of the space--time are closely related. To see this suppose we are in a gauge 
in which Dirac field is a constant. By applying a gauge transformation one can change 
it to a general space--time dependent function.
\begin{equation}
\beta=\beta_0 \longrightarrow \beta=\beta(x)=\beta_0\exp (-\Lambda(x))
\end{equation}
This gauge transformation is defined as:
\begin{equation}
\phi_\mu \longrightarrow \phi_\mu+\partial_\mu\Lambda
\end{equation}
So, the gauge in which the quantum mass is constant (and thus the quantum force is zero) 
and the gauge in which the quantum mass is space--time dependent are related to each other 
via a scale change. In other words, $\phi_\mu$ in the two gauges differ by 
$-\nabla_\mu(\beta/\beta_0)$. Since $\phi_\mu$ is a part of Weyl geometry, and Dirac field 
represents the quantum mass, one concludes that the quantum effects are geometrized.
One can see this fact also by referring to the equation (\ref{x13}) which shows that 
$\phi_\mu$ is not independent of $\beta$, so the Weyl vector is determined by quantum mass, 
and thus this geometrical aspect of the manifold is related to the quantum effects.
 In this way, the physical meaning of auxiliary Dirac field is clarified, as while as 
a suitable model for quantum gravity is introduced.

In a forthcoming paper we shall investigate the solutions of the
field equations of this theory.

\end{document}